\documentclass{aa}  %[referee]--colocar antes do aa
\usepackage{graphicx}
\usepackage{natbib}

\begin{document}
\title{On the influence of non-thermal pressure on the mass determination of galaxy clusters}
\author{T. F. Lagan\'{a} \inst{1}
\and R. S. de Souza\inst{1}
\and G. R. Keller\inst{1}}

  \offprints{Tatiana F. Lagan\'{a} \email{tflagana@astro.iag.usp.br}}

\institute{Universidade de S\~{a}o Paulo, Instituto de Astronomia, Geof\'{i}sica e Ci\^{e}ncias Atmosf\'{e}ricas, Departamento de Astronomia, Rua do Mat\~{a}o 1226, Cidade Universit\'{a}ria,
05508-090, S\~{a}o Paulo, SP, Brazil.}

\date{Received 17 February 2009 / Accepted 10 November 2009}

\authorrunning{Lagan\'{a}, Souza \& Keller}
   \titlerunning{Non-thermal pressure in clusters of galaxies}

\abstract
{}
{Given that in most cases just thermal pressure is taken into account in the hydrostatic equilibrium equation 
to estimate galaxy cluster mass,
the main purpose of this paper is to consider the contribution of all three non-thermal components 
to total mass measurements. 
The non-thermal pressure is composed by cosmic rays, turbulence and magnetic pressures.}
{To estimate the thermal pressure we used public XMM-\textit{Newton} archival data of five Abell clusters to derive temperature and density profiles.
To describe the magnetic pressure, we assume a radial distribution for the
magnetic field, $B(r) \propto \rho_{g}^{\alpha}$, to seek generality we assume $\alpha$ within the range of 0.5 to 0.9,
as indicated by observations and numerical simulations.
Turbulent motions and bulk velocities add a turbulent pressure, which is considered using an estimate from
numerical simulations.
For this component, we assume an isotropic pressure, $P_{\rm turb} = \frac{1}{3}\rho_{\rm g}(\sigma_{r}^{2}+\sigma_{t}^{2})$.
We also consider the contribution of cosmic ray pressure, $P_{cr}\propto r^{-0.5}$.
Thus, besides the gas (thermal) pressure, we include these three non-thermal components in the
magnetohydrostatic equilibrium equation and compare the total mass estimates with the values
obtained without them.}
{A consistent description for the non-thermal component could yield a variation in
mass estimates that extends from 10\% to $\sim$30\%. 
We verified that in the inner parts of cool core clusters the cosmic ray component is comparable to
the magnetic pressure, while in non-cool core clusters the cosmic ray component is dominant.
For cool core clusters the magnetic pressure is the dominant component, 
contributing more than 50\% of the total mass variation due to non-thermal pressure components. 
However, for non-cool core clusters,
the major influence comes from the cosmic ray pressure that accounts for more than 80\%
of the total mass variation due to non-thermal pressure effects. 
For our sample, the maximum influence of the turbulent component to the total mass variation can be almost 20\%.
Although all of the assumptions agree with previous works, 
it is important to notice that our results rely on the specific parametrization adopted in this work.
We show that this analysis can be regarded as a starting point for a more detailed and refined
exploration of the influence of non-thermal pressure in the intra-cluster medium (ICM).}
{}

\maketitle

\section{Introduction}
\label{intro}
Clusters of galaxies are powerful tools for investigations of cosmological interests. 
The evolution of the mass function of a cluster is highly sensitive to cosmological models since 
the matter density controls the rate at which structures grow \citep{Voit05}.
In order to use clusters of galaxies as observational probes 
of dark energy in the Universe and
to investigate the structure formation history including baryonic hydrodynamics, 
the non-thermal contribution must be well understood and quantified.

X-ray data are one of the most often methods used to determine the mass distribution of clusters of galaxies.
To do so, hydrostatic equilibrium is usually  assumed, and the observed gas density and temperature profiles are used
to compute the thermal pressure. 
In most  cases, only the gas (thermal) pressure is considered to evaluate the dynamical masses of galaxy clusters 
\citep[e.g.,][]{David95,WF95,Finoguenov01,Reip02}.
However, there is also a non-thermal pressure ($P_{\rm NT}$), composed by the magnetic ($P_{\rm B}$), 
turbulent ($P_{\rm turb}$) and  cosmic ray ($P_{\rm cr}$) components that is frequently assumed to be 
negligible and thus ignored. As a consequence of this, today the accuracy of the hydrostatic mass estimates 
is limited by the non-thermal 
pressure from these components. 

Despite the difficulty to reliably calculate the small-scale properties of the magnetic field,
the existence of intra-cluster magnetic fields is well established from the 
studies of the rotation measure of polarized
radio frequencies and synchrotron emission from diffuse sources 
\citep[e.g.,][]{Andernach88,giov93,Taylor94,Taylor02,Govoni04}.
More recently, another indication that the
intra-cluster medium (ICM) is permeated by a magnetic field came from the studies of X-ray cold fronts
\citep[sharp discontinuities in X-ray surface brightness profile and temperature,][]{Mark07}.
In these cases, a parallel magnetic field can suppress transport processes in the ICM,
making it difficult to mix different gas phases during a cluster merger.
Even a very weak magnetic field can effectively
inhibit transport processes such as thermal conduction and the settling of 
heavy ions \citep{Sarazin86,SS90}.

Strong magnetic fields can make a significant contribution to the gas pressure
support \citep{loeb94}, contributing with a non-thermal component in the magnetohydrostatic
equilibrium equation \citep{D01b,Vogt05}.
Indeed, magnetic fields as high as 10 - 100 $\mu$G were found in Hydra A \citep{Taylor93},
Cygnus A  \citep{Dreher87} and in 3C 295 \citep{Perley91}.
Moreover, \citet{Dolag99} performed numerical simulations and found that even clusters with an overall
small magnetic field can be penetrated partially by regions of high magnetic fields.

Although on average the magnetic pressure in simulations is
much smaller than the thermal pressure \citep[$\sim$ 5\%,][]{DS00}, there
are domains of high magnetic fields approaching or sometimes even
exceeding equipartition with the thermal energy. Previous studies \citep[][among others]{Dolag01a,Colafrancesco07}
have analyzed the effects of the magnetic pressure in simulated galaxy clusters.

Magnetic fields and turbulence are possibly related to one another.
It seems plausible that the turbulent motions
in the ICM can maintain the magnetic field by converting kinetic energy into magnetic energy \citep{Sanchez99}.
The observed small-scale turbulence in the ICM can be due to bulk velocities and ongoing merger of substructures. 
Gas turbulence on small scales can also be driven directly by motions of galaxies, 
as for instance by jets and bubbles from the active galactic nuclei \citep[AGN,][]{Churazov02}, 
although the latter may be confined to the inner regions of the cluster \citep{Lau09}. 
The presence of random gas motion can also contribute to the
pressure support in clusters of galaxies.

The chaotic nature of the ICM magnetic field makes
it difficult for energetic particles to scape from the cluster, and
thus cosmic-ray protons would be confined for timescales exceeding the
Hubble time. The electron cosmic rays, on the other hand, have
collisional and radiative lives much shorter. Thus, since the
ICM is permeated by significant magnetic fields, one would expect
the cosmic ray pressure to have some relevance in its support
against gravity.

To consider deviations from the standard assumptions in computing cluster total mass,
the main aim of this work is to analyze the effects of non-thermal pressure, that is 
to take into account magnetic,
turbulent and cosmic ray components. 
Hydrostatic masses were derived using X-ray observational data for 
five Abell clusters: A496, A2050, A1689, A2667 and A2631.
To do so, we use temperature and density profile fits from a previous work \citep{Lagana08}
and we introduce the $P_{\rm NT}$ contribution in the magnetohydrostatic equilibrium equation.
For these five clusters, we compare masses determined considering non-thermal pressure
($M_{\rm NTP} (r)$) with their hydrostatic values ($M (r)$).

The  paper is organized as follows. We show the data sample in Sect.~\ref{data}.
The non-thermal components are described in Sect.~\ref{NTC}. In this section
we describe the structure of the intra-cluster magnetic field, the 
turbulence in the ICM and the cosmic ray component. 
In Sect.~\ref{PNT}, we present the
method of determining  the cluster mass, including the effects of the $P_{NT}$. 
Our results, as well as a discussion of them are presented in Sect.~\ref{RD} and our  
conclusions in Sect.~\ref{conc}.

\section{Data sample}
\label{data}
The objects in our sample are within the redshift range of $ 0.03 < z < 0.3 $ and 
are drawn from a set of Abell clusters with available data in the XMM-Newton
public archive.
These clusters were previously analyzed by \citet{Lagana08}, 
who derived the density profiles fit parameters to
compute the total mass.
Although we have not used the object morphology as a criterion for the cluster selection, all these
clusters except A2631 have apparently symmetric X-ray isophotes, suggesting that they are relatively
relaxed. The deviations in the surface brightness profile of A2631, although clearly present, are not
very large and do not invalidate the assumption of spherical symmetry. However, we are aware of the fact
that it may affect total mass reconstruction, accounting for underestimated mass determinations
\citep{Piffaretti08}.

In Table~\ref{generalinfo}, we present the five Abell clusters used in this work,
specifying $r_{500}$,  the radius inside which the mean density exceeds the
critical density by a factor of 500.
All masses are computed inside $r_{500}$, as it is the largest radius for which the
current X-ray data require no model extrapolation
\citep{Vik06} and is about the virial radius \citep{LaceyCole93}.

\begin{table}
\centering
\caption{General cluster properties.}
\begin{tabular}{ccccc}
\hline\hline
 Cluster & R.A & DEC & ${\it z}$ &  $r_{500}$ \\
         & (J2000) & (J2000)&  & $h_{70}^{-1} kpc$\\
\hline\noalign{\smallskip}
A496  & 04 33 37.1 & -13 14 46 & 0.033   & 1480\\
A2050 & 15 16 21.6 & +00 05 59 & 0.1183  & 2172\\
A1689 & 13 11 34.2 & -01 21 56 & 0.1823  & 1785\\
A2667 & 23 51 47.1 & -26 00 18 & 0.23    & 2153\\
A2631 & 23 37 39.7 & +00 17 37 & 0.273  & 1976\\
\hline
\end{tabular}
\label{generalinfo}
\end{table}

Usually, the mass of a cluster  is  determined under the assumption of hydrostatic equilibrium without the
contribution of non-thermal pressure. In this case, the total mass relies on the temperature and density profiles.

Satellites with better spatial resolution (like the XMM-Newton and Chandra) showed a significant difference between the
surface brightness profile data points and the $\beta$-model \citep{CavFusco76,CavFusco78}
at small radii for cool core clusters \citep{JF84,XueWu2000}.
Based on this observational difference,
the $\beta$-model was used to describe
the density distribution of
non-cool core (NCC) clusters, while the S\'{e}rsic model \citep{Pislar97,Demarco03} was used to characterize
cool core (CC) clusters. For A2050 and A2631, the gas density ($\rho_{g}$) is described by
\begin{equation}
\label{roBeta}
\rho_{g}(r)=\rho_{0}\left(1+\frac{r^{2}}{r_{c}^2}\right)^{-3\beta/2},
\end{equation}
where $\rho_{0}$ and $r_{c}$ are the central gas density and the
gas core radius, respectively. The $\beta$ parameter determines the power-law
behavior at large radii.
For A496, A1689 and A2667 (CC clusters) the gas density profiles were fitted by the S\'ersic model given by
\begin{equation}
\label{roSersic}
\rho_{g}(r)=\rho_{0}\bigg(\frac{r}{a}\bigg)^{-p^{\prime}} \exp\left[-\left(\frac{r}{a}\right)^{\nu}\right],
\end{equation}
where $p^{\prime}=p/2$, $p=1-0.6097 \nu +0.05563 \nu^{2}$ and $a=a^{\prime}~2^{1/\nu}$ \citep{Durret05}.
The best-fit parameters were determined from the X-ray surface brightness profiles and were given by \citet{Lagana08}.

\section{Non-thermal components}
\label{NTC}

In this section we describe each non-thermal component
considered to contribute to the pressure support.

\subsection{The magnetic profile}
\label{profB}
In the 80s, \citet{Jaffe80} suggested that the intra-cluster magnetic field distribution should depend 
on the thermal gas density and on the distribution of massive galaxies, which means it would decline with cluster radius.
Cluster observations provided constraints on the radial gradient of the cluster magnetic field
\citep{Brunetti01,Govoni01,Feretti04}. The intensity of the magnetic field was found to decrease smoothly with
the cluster-centric radius, with a trend similar to that of the thermal gas.

From magnetohydrodynamic (MHD) cosmological simulations, an important characterization of the  cluster magnetic distribution was made by
\citet{Dolag99,Dolag02}. They studied the  correlation between X-ray surface brightness and Faraday rotation
measurements (FRMs) in clusters provided by X-ray and radio observations as well as from models for the ICM.
These authors performed cosmological MHD simulations in order to recover the correlation between these quantities. 
They found a relation between magnetic fields and the gas density of the cluster, suggesting 
that the cluster magnetic fields may span a wide range of spatial scales with a strength
that decreases with distance from the cluster center. \citet{murgia04} used numerical simulations to investigate 
the relation between magnetic fields and Faraday rotation effects in clusters. 
These latter authors compared their simulations with polarization properties of extended cluster 
radio sources in radio galaxies and halos. They considered that the intensity of magnetic fields
decreases from the cluster center in agreement with previous results \citep{Dolag99,Dolag02}.

\citet{Dolag99} found that the observed intra-cluster magnetic field can be reproduced
by the evolution of an initial magnetic field at redshift 15 that was amplified by compression during the
cluster collapse. One of their important results was that the intra-cluster magnetic field strength is
proportional to the gas density at any point ($B(r) \propto \rho (r)$).

\citet{Colafrancesco07} studied the influence of magnetic fields on the  main structural
properties of virialized groups and clusters, assuming that it scales with a 
density of $B(r) \propto \rho^{\alpha}(r)$, as previously proposed. The same power law
dependence on the density was used by \citet{Zhang04} and \citet{Koch03} to estimate the effect of the
intra-cluster magnetic field on the Sunyaev-Zel'dovich power spectrum.

Motivated by these previous works, we assumed a parametric
form for the radial distribution of the magnetic field

\begin{equation}
\label{eqB}
B(r)= B_{0}\left(\frac{\rho_{g}(r)}{\rho_{0}}\right)^{\alpha},
\end{equation}
where $B_{0}$ is the central value of the magnetic field and $\alpha$ is the shape parameter.
Unfortunately, there are no measurements for the magnetic profile for any of the
clusters in our sample. Thus, we have to use results from the literature to constrain
$B_{0}$ and $\alpha$.

The effective strength and structure of these magnetic fields provide the main challenge,
because different methods of analysis give different values for magnetic strength. 
An estimated consideration of the equipartition of the magnetic field strength averaged
over the entire halo volume gives magnetic field strengths of $\sim 0.1-1 \mu G$ 
\citep[][and references therein]{govoni01a, murgia04}.

\citet{Feretti99} estimated
that the magnetic field in the ICM of A119 should range between $5-10 \mu G$. 
\citet{Bagchi98} found $B \approx 1\mu$G for the cluster-scale
magnetic field strength.
In a more recent work, \citet{Clarke01} studied a sample of 16 ``normal'' low-redshift (z $<$ 0.1) galaxy clusters,
finding that the ICM is permeated with magnetic fields at levels of 4-8 $\mu G$.  \citet{Taylor93} found higher central values,
$B \sim 6-30 \mu$G, for the ICM magnetic fields. \citet{Allen01} claimed that the central value of the
magnetic fields can be $B=12\mu$G.
FMRs of radio sources provide
magnetic fields of $\sim 5-30 \mu G$ in cooling flow clusters 
(e.g., 3C 295 \citep{Allen01}, Coma \citep{Feretti95} and Hydra A \citep{Taylor93}) 
where extremely high FRMs have been revealed.
\citet{Carilli02} affirmed that its strength in the center of cooling-core clusters
can reach levels of $10-40 \mu $G.

On the other hand, lower magnetic fields $(\sim 2-8 \mu G)$ have also been detected
in clusters without cooling flows \citep[e.g.,][]{Feretti99a, Taylor01, Eilek02}. 

The magnetic field strengths obtained from FRMs arguments
are higher than the values derived either
from the radio data or from inverse Compton X-ray
emission. The  values deduced from radio
synchrotron emission and from inverse Compton refer
to averages over large volumes. Instead, FRMs estimates give
a weighted average of the field and gas density along the
line of sight and could be sensitive to the presence of filamentary
structure in the cluster. They could
therefore be higher than the average cluster value.
However, as
pointed out by \citet{Carilli02},
all of these techniques are based on several assumptions.
For example, the observed FRMs have been interpreted
until now in terms of simple analytical models which consider
single-scale magnetic fields, while equipartition calculations in
radio halos assume spatially uniform magnetic fields.

There are not many works that studied the power spectrum of the intra-cluster
magnetic field fluctuations. However, \citet{Ensslin03} and \citet{Vogt03, Vogt05} by using a new 
semi-analytical
technique showed that, for those cluster sources for
which a very detailed FRM image is available, the magnetic field
power spectrum can be estimated.

To take into account all these observational results, we considered conservative values in our analysis, 
adopting for cooling core clusters 
values for $B_{0}$ ranging from a low value of 5$\mu$G up to a strong value of 30$\mu$G. 
But for non-cooling core clusters
we considered the central strength of the magnetic field to vary from 
2$\mu$G $<\rm B_{0}<$ 8$\mu$G. 
The values adopted agree with both observations and with numerical 
simulations that considered higher values for magnetic fields in the core of the cluster and a
decrease towards the outskirts.

If the ratio of magnetic and thermal pressure is constant throughout the cluster, then $\alpha = 0.5$, 
if we have an homogeneous seed magnetic field compressed during cluster formation, one expects $\alpha = 2/3$.  
\citet{Vogt05} performed a Bayesian maximum likelihood analysis of Faraday rotation measure in order to derive a power 
spectrum of cluster magnetic fields. In their analysis they used three different values for $\alpha = 0.1, 0.5, 1.0$. 
They concluded that values of $\alpha = 1.0$ are unlikely, but  models with $\alpha = 0.1-0.5$ match very well with their calculations. 
From observational results, \citet{D01b} found $\alpha=0.9$ for A119 and $\alpha=0.5$ for 3C 129. 
If one combines the measurements of the four clusters the data analysis performed by these latter authors leads
to a slightly lower slope, of this correlation ($\alpha=0.8$).
Therefore, to be conservative, we decide to vary the shape parameter within the interval of $ 0.5 < \alpha < 0.9$ 
to take into account all the above results.

\begin{figure}
\centering 
\includegraphics[width=0.3\textwidth]{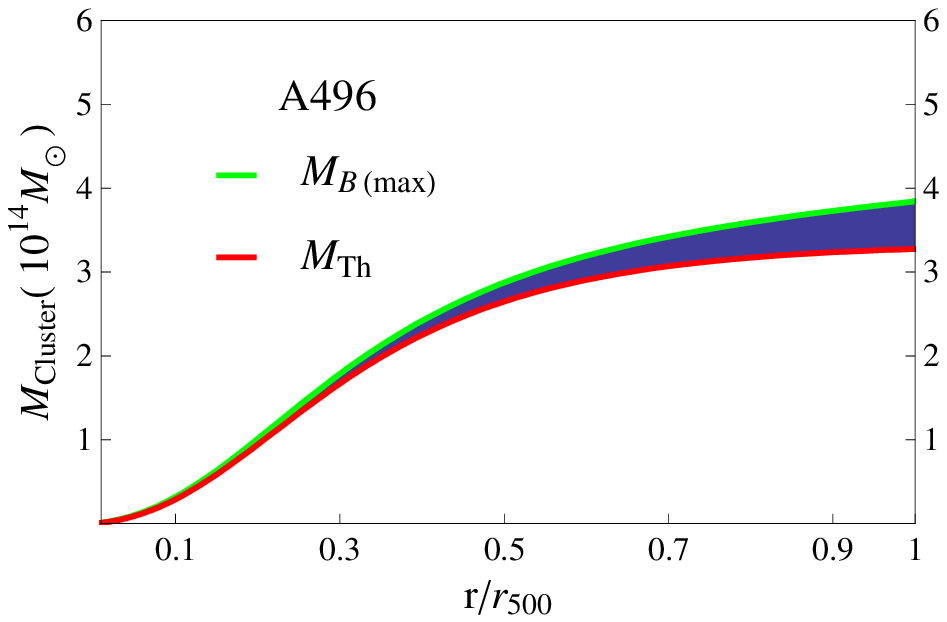}
\includegraphics[width=0.3\textwidth]{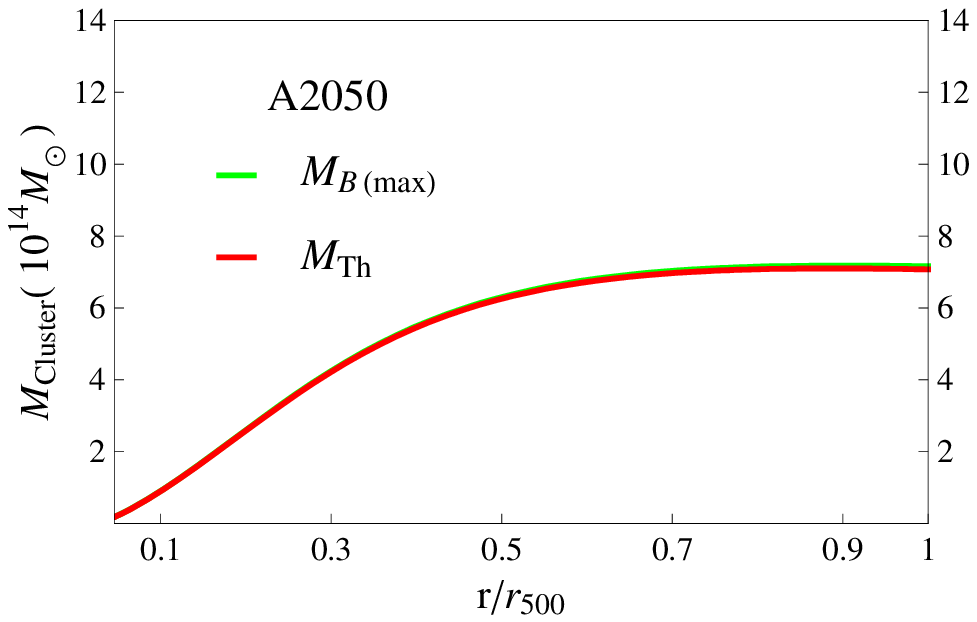}
\includegraphics[width=0.3\textwidth]{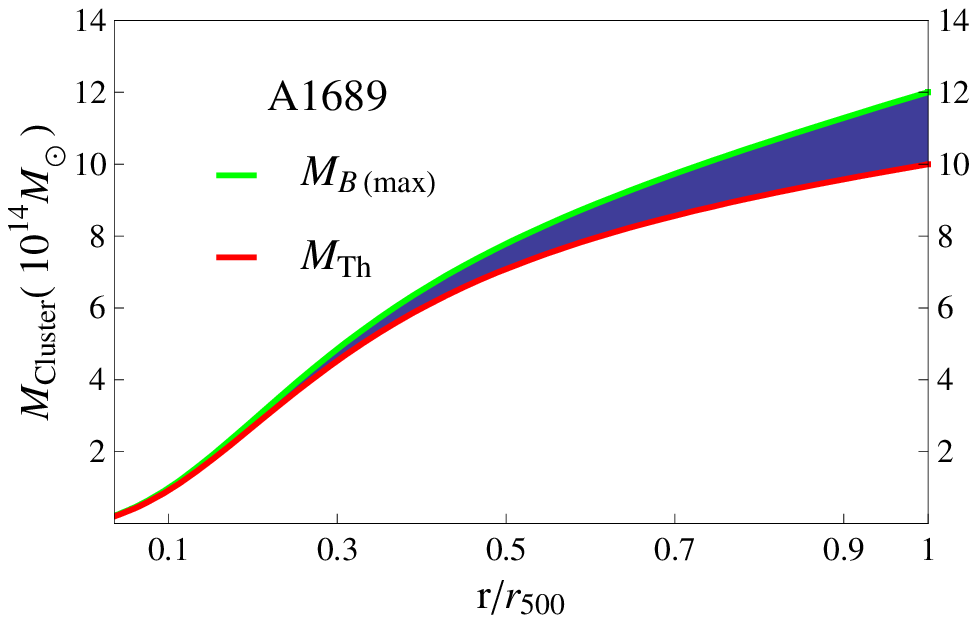}
\includegraphics[width=0.3\textwidth]{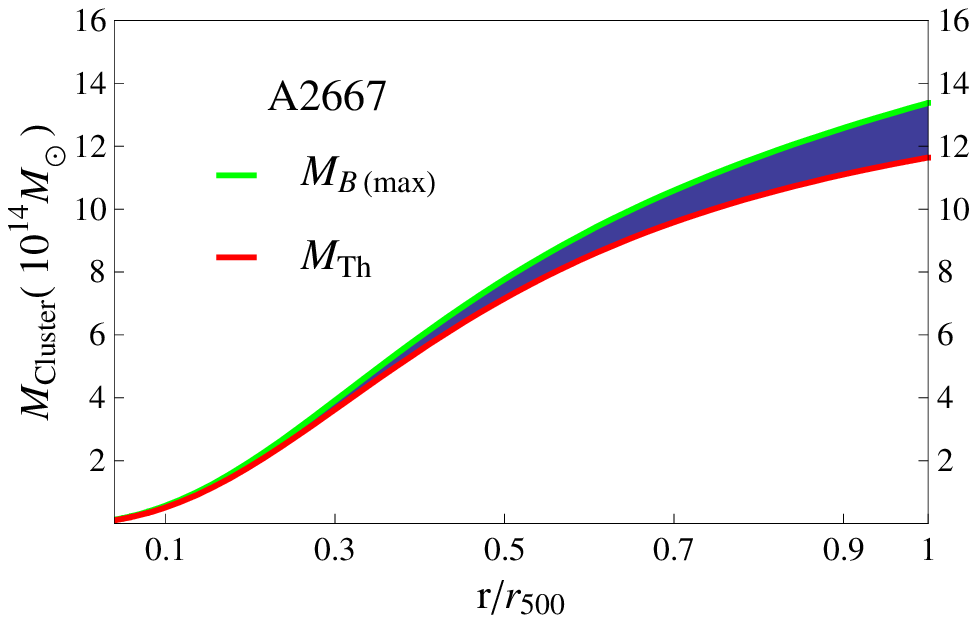}
\includegraphics[width=0.3\textwidth]{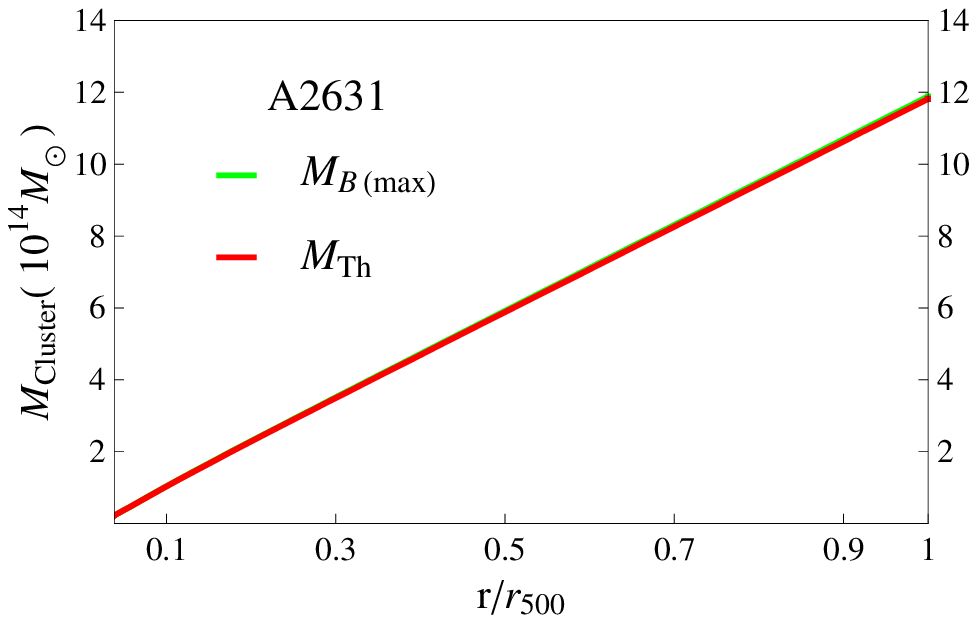}
\caption{\small Variation in the mass profile due to the magnetic pressure as the only non-thermal component for the clusters 
in our sample. The red lines represent the hydrostatic mass profiles, while the green lines show the maximum variation
of mass profiles due to the contribution of the magnetic pressure. The blue zone represents the region of mass profile variation
depending on the central value strength ($B_{0}$) and the shape parameter $\alpha$.}
\label{FigCampoB}
\end{figure}

In Fig.~\ref{FigCampoB} we show the mass profile variation due to the magnetic pressure as 
the only non-thermal component compared to the hydrostatic mass profile. 
From this Figure we see that the magnetohydrostatic profiles for NCC clusters (A2050 and A12631) 
present little difference when compared to the hydrostatic profile. This little influence on 
the magnetic pressure in NCC cluster is due to the central value assumed for these clusters.
As the central strength of the magnetic field is lower in NCC clusters, the influence of the
magnetic pressure is also lower compared to a CC cluster. 
For CC clusters we note that the difference between the hydrostatic and the magnetohydrostatic 
profiles becomes more pronounced for large radii, that is for $ r > 0.5~r_{500}$.

\citet{Churazov08} measured the contribution of the non-thermal pressure in two early-type galaxies
that reside in the center of two nearby cool core clusters (Virgo and Fornax). 
With a similar approach, these authors considered the
contribution of all non-thermal components in the same system, combining the contribution of cosmic rays, 
magnetic fields and microturbulence to
the total pressure. They suggested that the $P_{NT}$ component can account for 10\% of the gas thermal pressure
in the core of these galaxies (NGC 1399 and M87). Assuming that the magnetic component is 
the only non-thermal pressure, \citet{Churazov08} constrained the upper limits on 
magnetic field to be $\sim$ 20-30 $\mu$G.
Besides, \citet{Ajello09} reported the detection of ten merging-clusters
in the 15-55 keV energy band using \textit{Swift's Burst Alert Telescope}
(BAT\footnote{\tiny http://swift.gsfc.nasa.gov/docs/swift/about\_swift/bat\_desc.html}).
These authors coupled radio synchrotron emission (inverse Compton) with X-ray data,
putting the constraints on the lower limit of the magnetic field to be typically between 0.1-0.5$\mu$G.
We can see from these Papers mentioned above that both the upper and lower limits adopted in this work 
are in accordance with the recent literature.

\subsection{Turbulence in clusters of galaxies}
\label{Turbsection}

With the advent of high-resolution observations from the \textit{Chandra} and XMM-\textit{Newton} satellites,
temperature maps of the X-ray emitting gas have shown that even apparently relaxed clusters could have small substructures
as imprints of recent minor-mergers  \citep[e.g.,][]{Finoguenov05,Durret08}.

It is widely accepted that the ICM is probably
turbulent, and mergers of galactic sub-clusters may be one of
the major energy injection mechanisms \citep[see][and references therein]
{Sarazin02,Brunetti03,Lazarian06}. 
Random gas motions can also maintain and amplify cluster magnetic fields via
dynamo processes \citep{Roettiger99,Subramanian06} and contribute to 
the acceleration of cosmic rays in the ICM \citep{Brunetti07}.

Generally the  models assume a scenario with the
scale  for the injection of energy  of 100-500 kpc and
the injection velocity of the order of $10^{3}$ km/s.

Since the rate of dissipation of the turbulent energy cannot
exceed the X-ray luminosity of the cluster ($L_X$) in a steady-state, i.e., $\frac12 v_0^3/l_0\la
L_X/M_\mathrm{g}$ , where $v_0$ and $l_0$ are the turbulent speed and the scale respectively, 
and $M_\mathrm{g}$ is the  gas mass,
we have an upper limit on the turbulent velocity as follows \citep{Subramanian06}

\begin{equation}
v_0\la 180\,\frac{\mathrm{km}}{\mathrm{s}}
\left(\frac{l_{0}}{200 \rm kpc}\right)^\frac{1}{3}\!\!
\left(\frac{L_{X}}{10^{45} \rm erg/\mathrm{s}}\right)^\frac{1}{3}\!\!
\left(\frac{M_\mathrm{g}}{10^{14}M_\odot}\right)^\frac{1}{3}\!\!.
\end{equation}

\citet{Norman99} found that the
ICM becomes turbulent during cluster formation, with turbulent
velocities of about $400~\rm km/s$ within $1~\rm Mpc$ from the center of a cluster and
eddy sizes ranging from 50 to 500 kpc. In the cluster merger model
of \citet{Ricker01}, they found large-scale turbulence  with eddy sizes up to
several hundred kiloparsecs and
turbulent velocities of $\sim 100 - 400~\rm km /s$.

From the analysis of
pressure fluctuations as revealed in X-ray observations, \citet{Schuecker04}
argue that the integral turbulent scale in the Coma cluster is close to
100 kpc, and they assume a turbulent speed of $250~\rm km/s$ at that scale.

Although a number of other studies have examined random gas motions 
and their effect on the mass estimate \citep[e.g.,][]{Rasia04,Dolag05,Rasia06}, 
most of them have used simulations
with SPH gas dynamics. The magnitude and effects of gas motions
in such simulations depends on the specific treatment of
artificial viscosity \citep{Dolag05}. 
Thus, we used the results from \citet{Lau09}, 
which employed simulations with Eulerian gas dynamics with very low
numerical viscosity. This approach is therefore useful in evaluating possible
differences between numerical techniques and systematic theoretical
uncertainties.

In order to quantify the importance of pressure support from random gas motions in clusters,
we can write the following relation for the isotropic turbulent pressure $P_{\rm turb}$ \citep{Lau09}:
\begin{equation}
P_{\rm turb} = \frac{1}{3}\rho_{\rm g}(\sigma_{r}^{2}+\sigma_{t}^{2}),
\end{equation}
where $\sigma_{r}$ and  $\sigma_{t}$ are the radial and tangential  dispersion velocity of the intra-cluster gas respectively. 
For 16 simulated clusters with virial masses within the range of $(5 \times 10^{13} - 2 \times 10^{15})~ \rm M_{\odot}$, \citet{Lau09} found that 
gas motions contribute up to $\sim 5\%-15\%$ of the total pressure support in relaxed clusters.
Thus, on average the total mass estimate is biased low by about $8 \pm 2$\%
(at $r_{500}$) in relaxed systems and $11 \pm 6$\% in unrelaxed systems.
These results agree with previous studies 
\citep{Evrard90,Rasia04,Nagai07,Piffaretti08}, 
with contributions increasing along the radius.

Note that random gas pressure and its
gradient is sensitive to small-scale clumps and any pressure
inhomogeneity, and these sources could potentially bias the
measurements of the pressure gradient and hence the hydrostatic
mass estimate. In order to minimize this bias 
\citet{Lau09} removed subhaloes with a mass greater than $10^{12} h^{-1} ~M_{\odot}$ and the mass
within their tidal radius from their calculation (see their paper for further details).

\begin{figure}
\centering
\includegraphics[width=0.3\textwidth]{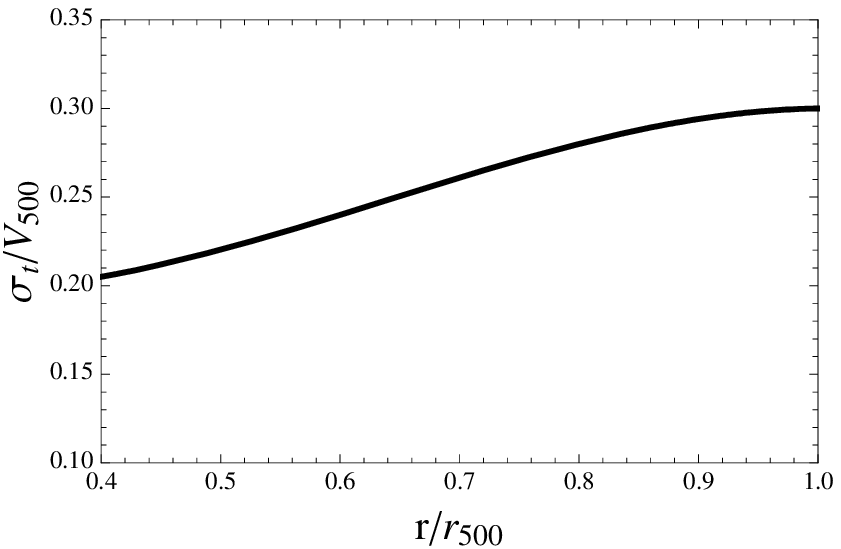}
\includegraphics[width=0.3\textwidth]{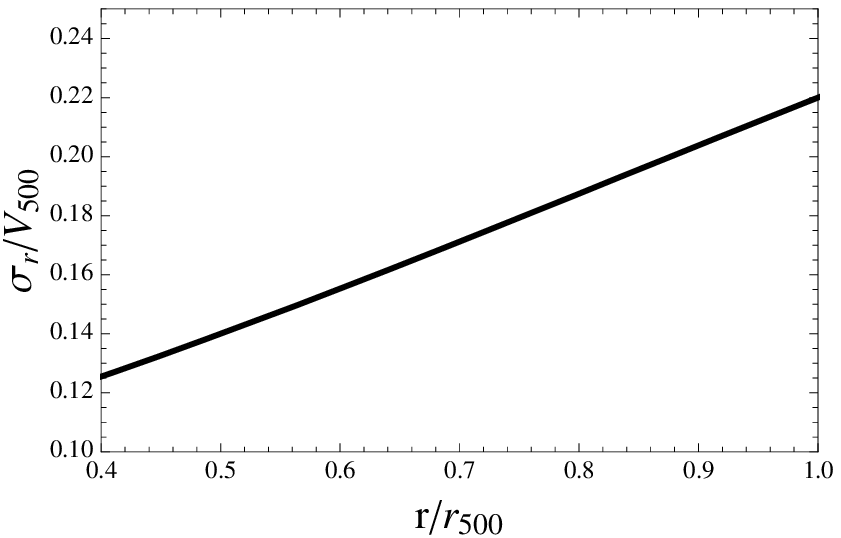}
\caption{\small Velocity dispersion profiles derived from numerical simulation of \citet{Lau09}.
Upper panel: tangential velocity profile used to estimate the turbulent pressure. 
Lower panel: radial velocity profile used to estimate the turbulent pressure.}
\label{VPs}
\end{figure}

In our analysis, we used radial and tangential velocity dispersion profiles (see Fig.\ref{VPs}) 
based on the numerical simulation of \citet{Lau09}. 
We show these profiles in Fig.~\ref{VPs} where the velocity
\begin{equation}
V_{500} = \sqrt{\frac{GM}{r_{500}}}
\end{equation}
is the circular velocity at $r_{500}$.

\begin{figure}
\centering
\includegraphics[width=0.3\textwidth]{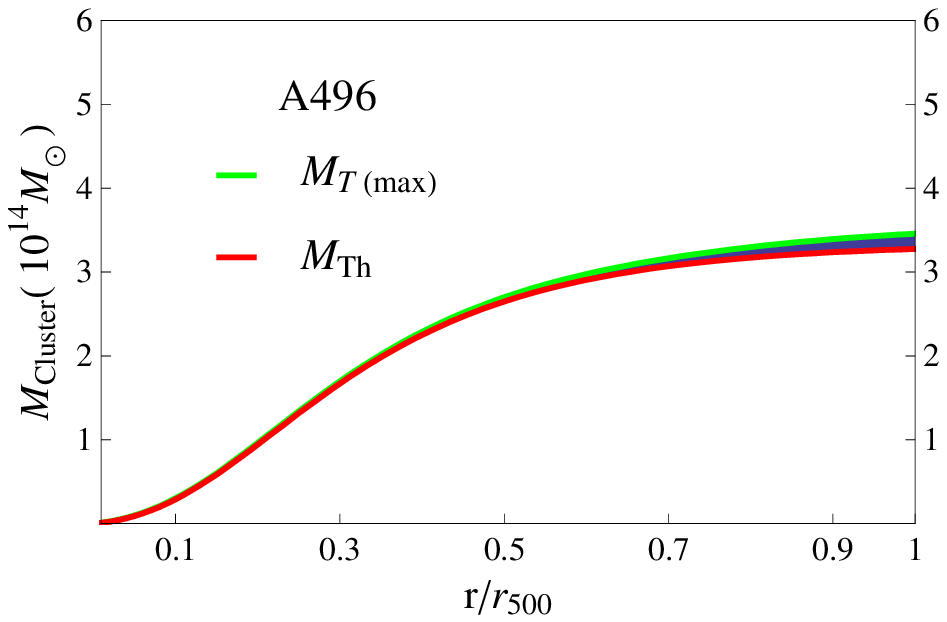}
\includegraphics[width=0.3\textwidth]{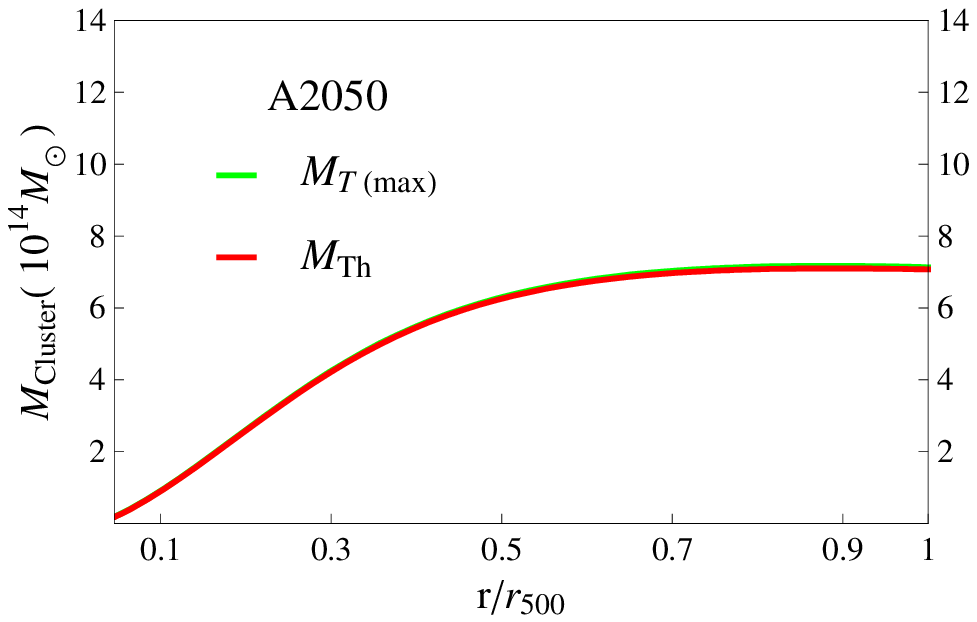}
\includegraphics[width=0.3\textwidth]{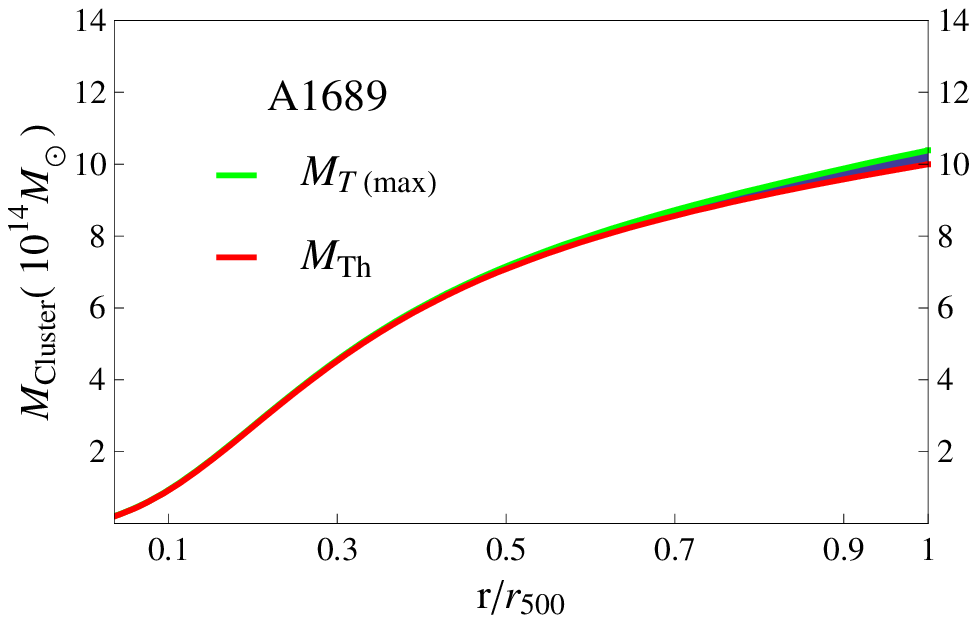}
\includegraphics[width=0.3\textwidth]{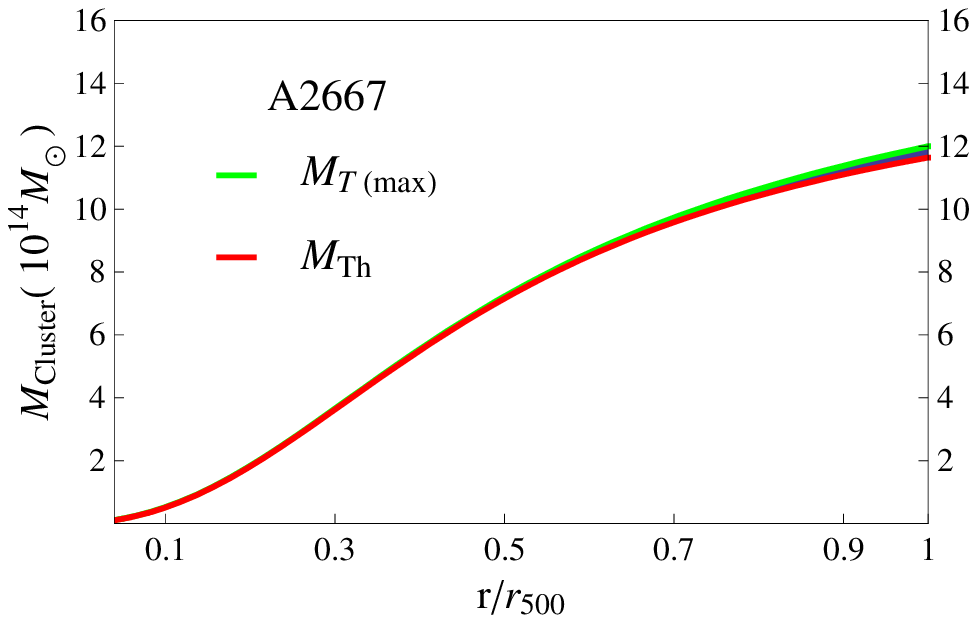}
\includegraphics[width=0.3\textwidth]{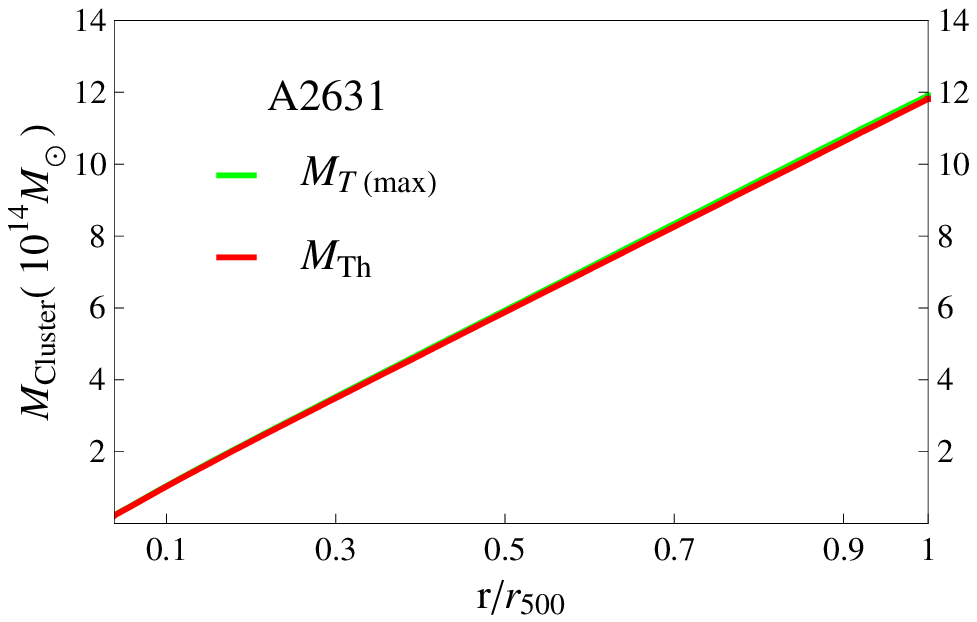}
\caption{\small Variation in the mass profile due to the turbulent pressure as the only 
non-thermal component for the clusters 
in our sample. The red line represents the hydrostatic mass profile, 
while the green line shows the maximum variation
of the mass profile due to the contribution of the turbulent pressure. 
The blue zone represents the region of the mass profile variation
depending on the radial and tangential dispersion velocities.}
\label{FigTurb}
\end{figure}

In Fig.~\ref{FigTurb} we show the mass profile variation due to the turbulent pressure as the only non-thermal component, 
compared to the hydrostatic mass profile. From this Figure we see that the influence of the turbulent pressure in 
the mass estimates is small regardless of weather the cluster is a non-cool core or cool core cluster.
Our sample has masses within the range of $ (0.3 -12)\times ~10^{14} \rm M_{\odot}$,
and we found that our mass estimates can be biased low by about $\sim$5\%.

\subsection{Cosmic ray pressure}
\label{CRsection}

Cosmic ray protons can play an important role within ICM contributing to the 
equilibrium on the pressure support. Cosmic ray protons (CRp) and cosmic ray electrons (CRe) 
can be injected into the ICM by 
three different processes which produce morphologically different radio signature \citep{Brunetti03,Pfrommer04}: 
\begin{itemize}
\item{shock acceleration: natural acceleration mechanism such as structure formation and merger shocks 
\citep{Ptuskin09,Ostrowski02}}
\item{reaccelerated electrons: reaccelerated relativistic particles injected by 
sources like radio galaxies, supernova remnants, merger shocks, galactic winds, etc. 
\citep{Ensslin97,Berezinsky97,Pfrommer04}.}
\item{particles of hadronic origin: CRp can eventually interact with the thermal ambient gas
producing secondary electrons, neutrinos and $\gamma$-rays in an inelastic collision which generates a 
radio halo through synchrotron emission.}
\end{itemize}

Evidence of non-thermal electrons in the ICM exist in the form of
synchrotron radio emission \citep{Ferretti08,Brunetti08}, excess of extreme-ultraviolet (EUV) and
hard X-ray radiation \citep{Bowyer98,Lieu99,FF04}. Another consequence of the presence of 
cosmic rays in the ICM is the production of gamma rays.

In order to consider the contribution of cosmic ray pressure to the ICM,
we followed the prescription of \citet{Ando08}.
We can define a relative contribution of the cosmic ray pressure support as
\begin{equation}
Y_{p} \equiv \frac{P_{\rm cr}}{P_{\rm g}},
\end{equation}
where this ratio can be parametrized using a  power law
\begin{equation}
Y_{p}(r) = Y_{p0}\left(\frac{r}{r_{0}}\right)^{\Psi},
\end{equation}
where the subscript 0 represents values in the central region. \citet{Sijacki08}, using numerical
simulations, followed  the evolution of the cosmic-ray  plasma inside the bubbles, considering both
its hydrodynamical interactions and the dissipation processes relevant to the cosmic ray  population.  
They found that cosmic ray pressure is most relevant in the center of the clusters, being comparable
to the thermal pressure for $r < 50 \rm h^{-1} \rm kpc$.  

Recent studies have highlighted that CRs can be dynamically important in galaxy clusters because
they put constraints on the fraction of cosmic-ray pressure with respect to the thermal pressure ($Y_{p}$). 
Since the \textit{Energetic Gamma Ray Experiment Telescope} 
(EGRET\footnote{http://heasarc.gsfc.nasa.gov/docs/cgro/cossc/egret/})
did not detect $\gamma$-ray emission from clusters in the GeV band \citep{Reimer03},
constraints on the fraction of cosmic ray pressure have been placed in cosmological simulations
of the large scale structure. 
In nearby rich clusters, this component should amount to about $\sim$10-26\%
of thermal pressure \citep{Ensslin97,Miniati01,Miniati03}.
By comparing the integrated $\gamma$-ray flux above 100 MeV to EGRET upper limits, 
\citet{Pfrommer04} constrained the CRp scaling parameter in their simulation
of nearby cooling-flow clusters. Thus they were able to infer that the $P_{\rm cr}$ 
accounts for less than 30\% of the thermal pressure. But
\citet{Sijacki08} affirmed that this component can reach up to 50\% of the
central gas pressure in clusters.

Despite all the effort in computing the cosmic ray pressure, 
the distribution of cosmic-rays in ICM is yet poorly known, and direct 
evidence for cosmic-ray ions in the ICM is still lacking.
The measurements cited above indicate that cosmic ray pressure accounts for a minor
contribution to the dynamical support \citep{Ando08}.
We expect that future experiments like the \textit{Imaging Air Cherenkov Telescopes}
(IACTs\footnote{http://magic.mppmu.mpg.de/introduction/iact.html}, 
which will work in the TeV band) and the
\textit{Gamma-ray Large Area Space Telescope} (GLAST\footnote{http://www-glast.stanford.edu/},
which will work in the GeV band) will be able to provide better constraints to $P_{cr}$ in clusters. 

Bearing these results in mind, we used central values for the ratio between the 
cosmic ray pressure and the thermal pressure between 10\%-50\%. 
To which end, we adopted $0.1 < Y_{p0} < 0.5$.

The value of $\Psi$ depends on the model adopted
for cosmic ray dynamics in clusters. In the simplest model $\Psi = 0$, the energy distribution of
cosmic rays follows precisely the thermal gas in the cluster.
The recent radiative simulations performed by \citet{Pfrommer07} showed that $\Psi$ takes a value of
-0.5. We considered $\Psi = -0.5$ based in \citep{Pfrommer07}. 

\begin{figure}
\centering
\includegraphics[width=0.3\textwidth]{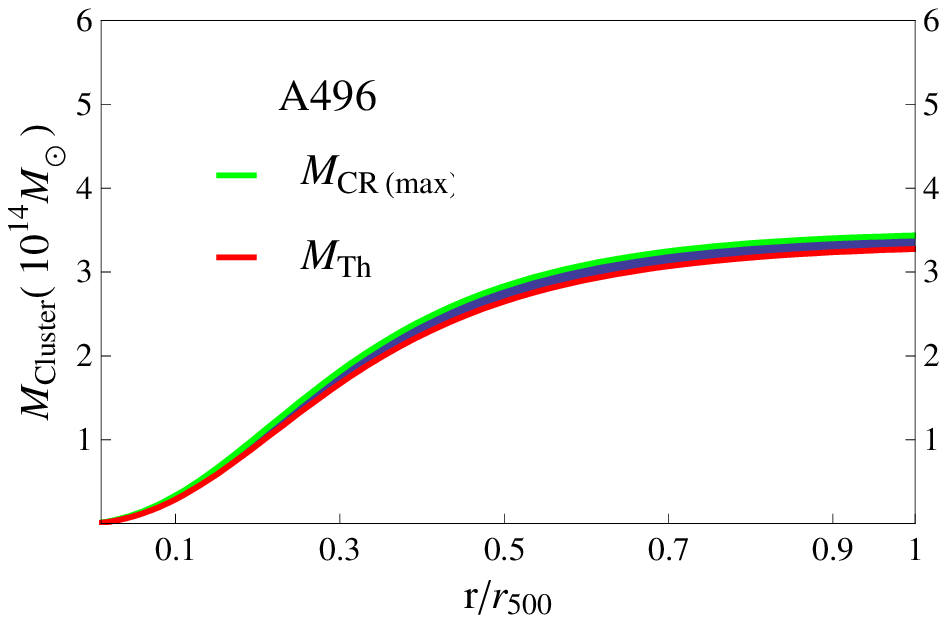}
\includegraphics[width=0.3\textwidth]{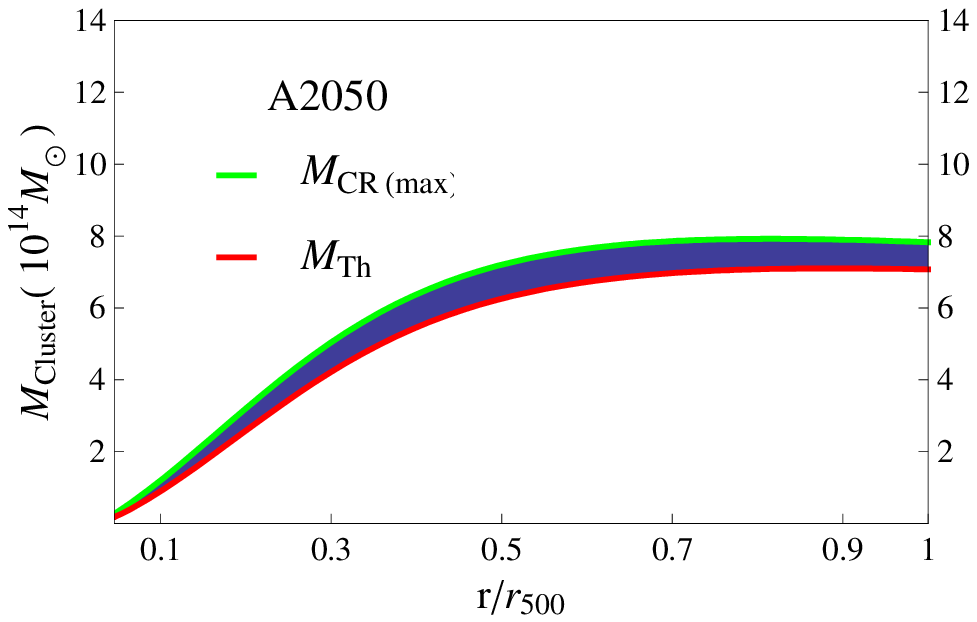}
\includegraphics[width=0.3\textwidth]{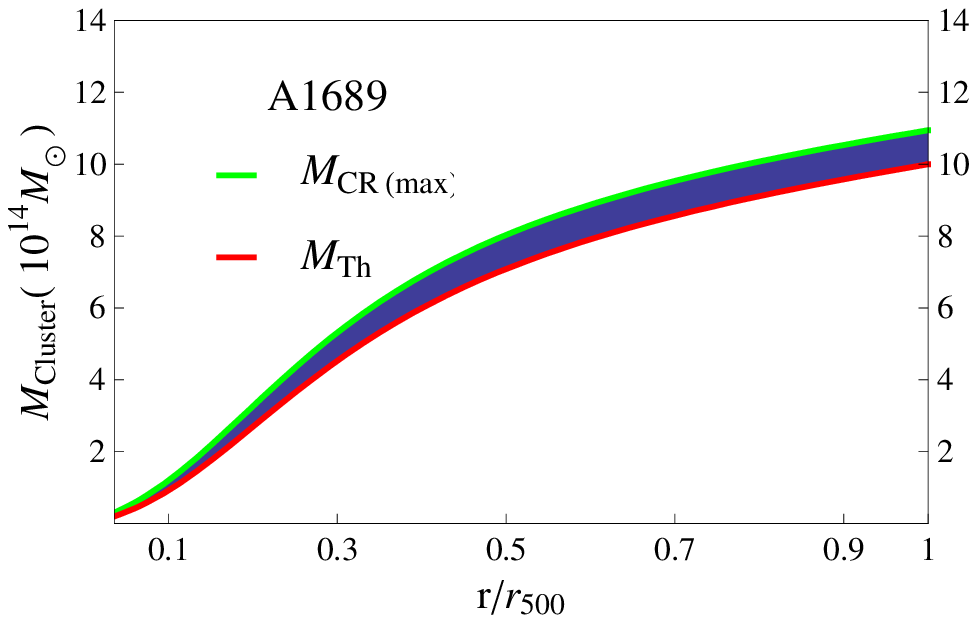}
\includegraphics[width=0.3\textwidth]{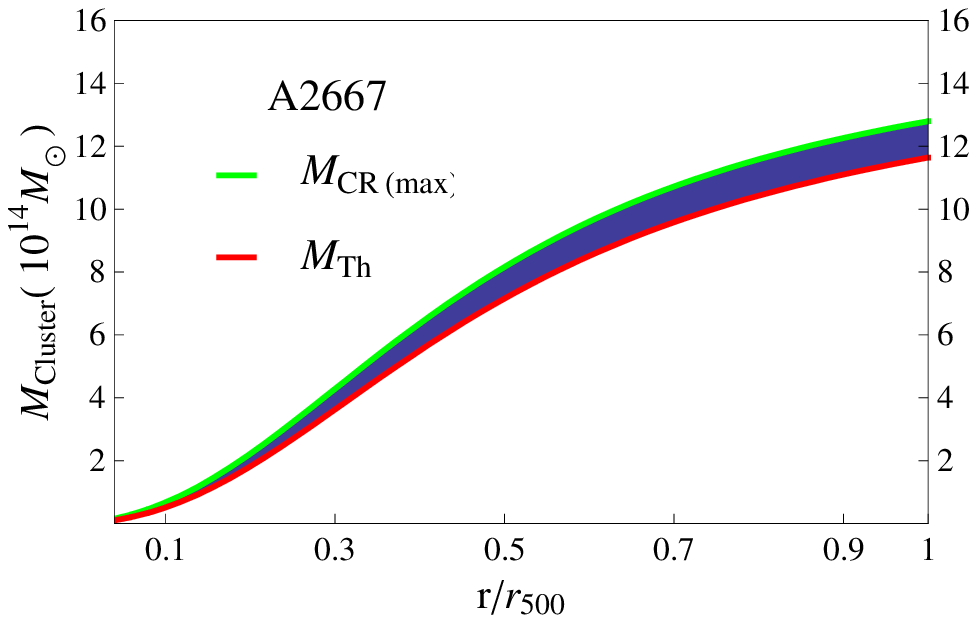}
\includegraphics[width=0.3\textwidth]{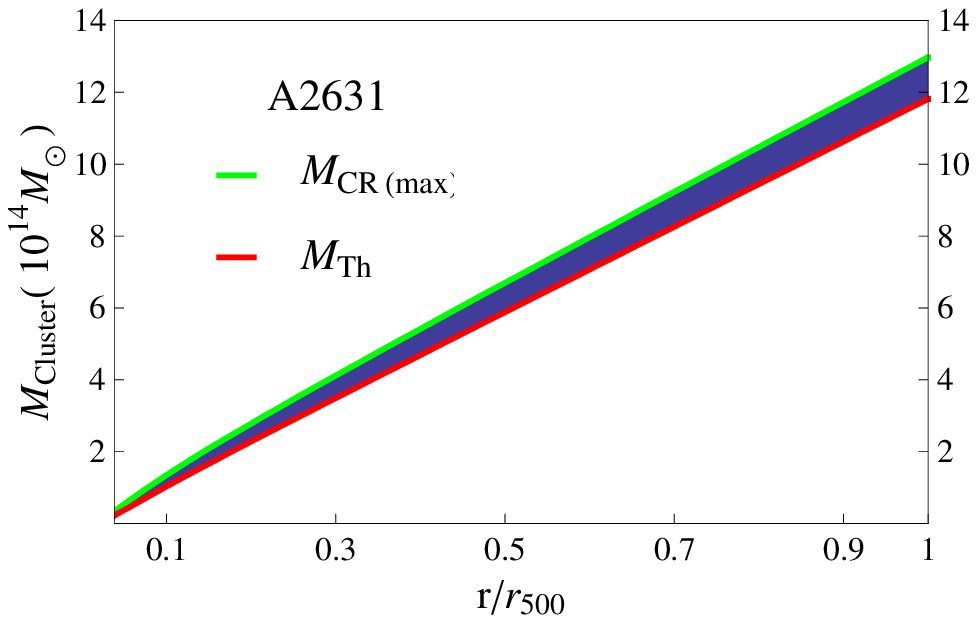}
\caption{\small Variation in the mass profile due to the cosmic ray pressure as the only non-thermal
component for the clusters
in our sample. The red line represents the hydrostatic mass profile, while the green line shows the
maximum variation
of the mass profile due to the contribution of the magnetic pressure. The blue zone represents the
region of the mass profile variation
depending on the $\Psi$ parameter.}
\label{FigRC}
\end{figure}

In Fig.~\ref{FigRC} we show the mass profile variation due to the cosmic ray pressure as the only
non-thermal component, compared to the hydrostatic mass profile.
Comparing the results presented in Fig.~\ref{FigCampoB} with Fig.~\ref{FigRC} 
we verify that the cosmic ray pressure is the most important non-thermal component 
for NCC clusters (see also Table~\ref{TabSigmaMass}).

\section{Mass determination including the effects of non-thermal pressure}
\label{PNT}

We have assumed that the cluster is spherically symmetric, the gas is in magnetohydrostatic equilibrium and 
that consequently, the gas pressure and the non-thermal components support the ICM against gravity.
We can then  write:
\begin{equation}
\frac{d(P_{\rm g} + P_{\rm B} + P_{\rm turb}+P_{cr})}{dr} = - \rho_{g} \frac{G M_{\rm NTP}(r)}{r^{2}},
\end{equation}
where  the gas pressure ($P_{\rm g}$)  at a temperature T is $ \rm \rho_{\rm g} kT / \mu m_{p}$,
the magnetic pressure ($P_{\rm B}$) is $\langle B^{2}\rangle /8 \pi$,
the turbulent pressure ($P_{\rm turb}$) is  $\frac{1}{3}\rho_{\rm g}(\sigma_{r}^{2}+\sigma_{t}^{2})$, 
cosmic ray pressure ($P_{CR}$) is $Y_{p0} P_{g}\left(r/r_{0}\right)^{\Psi}$,
$G$ is the gravitational constant and $M_{\rm NTP}$ is the total mass inside a radius $r$.
In our case, we computed the total masses within $r_{\rm 500}$.
Considering the influence of non-thermal pressures  
in the magnetohydrostatic
equilibrium equation, we have the following expression for the total mass of the cluster:
\begin{eqnarray}
\lefteqn{M_{\rm PNT}(r) = - \frac{k_{B}T(r)}{G\mu m_{H}}r\left(\frac{\mathrm{d} \ln \rho_{g}(r)}{\mathrm{d} \ln r} + \frac{\mathrm{d} \ln T(r)}{\mathrm{d} \ln r}\right)}
\nonumber \\
&&{} - \frac{r^{2}}{8 \pi \rho_{g}(r) G}\frac{\mathrm{d} B(r)^2}{d r} - \frac{r^{2}}{2 \rho_{g}(r) G} \frac{\mathrm{d}}{\mathrm{d}r} ({\rho_{g}(r) \sigma_{r}^{2}(r)})\nonumber \\&&{} -\frac{r}{G}(2\sigma_{r}^{2}(r)-\sigma_{t}^{2}(r))-\frac{r^{2}}{G\rho_{g}(r)}\frac{dP_{cr}(r)}{dr},
\end{eqnarray}
where $\mu$ is the mean molecular weight, $m_{H}$ is the hydrogen mass, $k$ is the Boltzmann
constant, $T(r)$ is the temperature profile, $B(r)$ is the magnetic profile described in
Sect.~\ref{profB}, and $\sigma_{r}$ and $\sigma_{t}$ are the radial and tangential 
dispersion velocity of the intra-cluster gas, respectively (see Sect. \ref{Turbsection}).

\section{Results and discussion}
\label{RD}

\begin{figure*}[h!]
\centering
\includegraphics[width=0.34\textwidth]{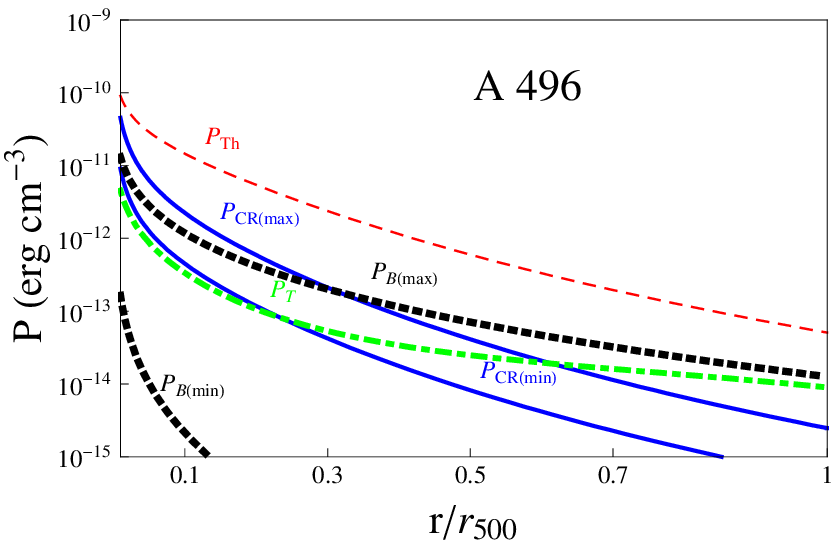}
\includegraphics[width=0.34\textwidth]{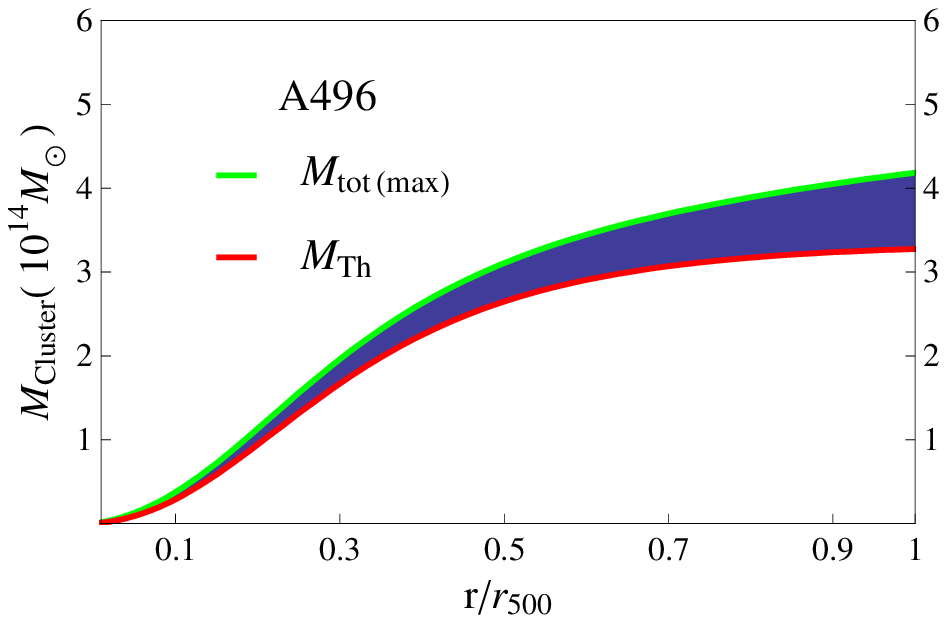}
\includegraphics[width=0.34\textwidth]{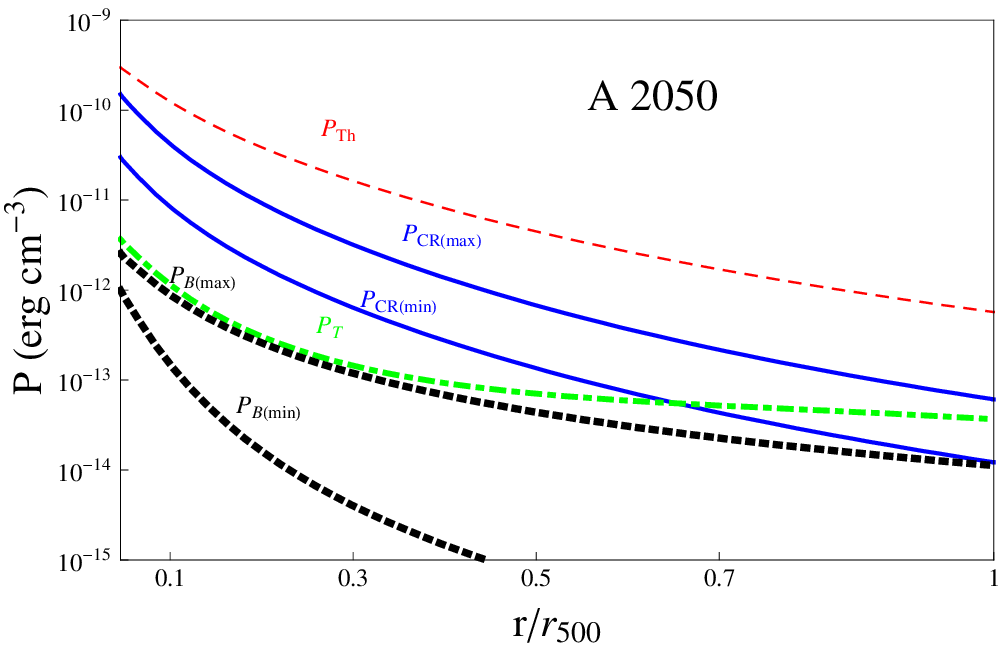}
\includegraphics[width=0.34\textwidth]{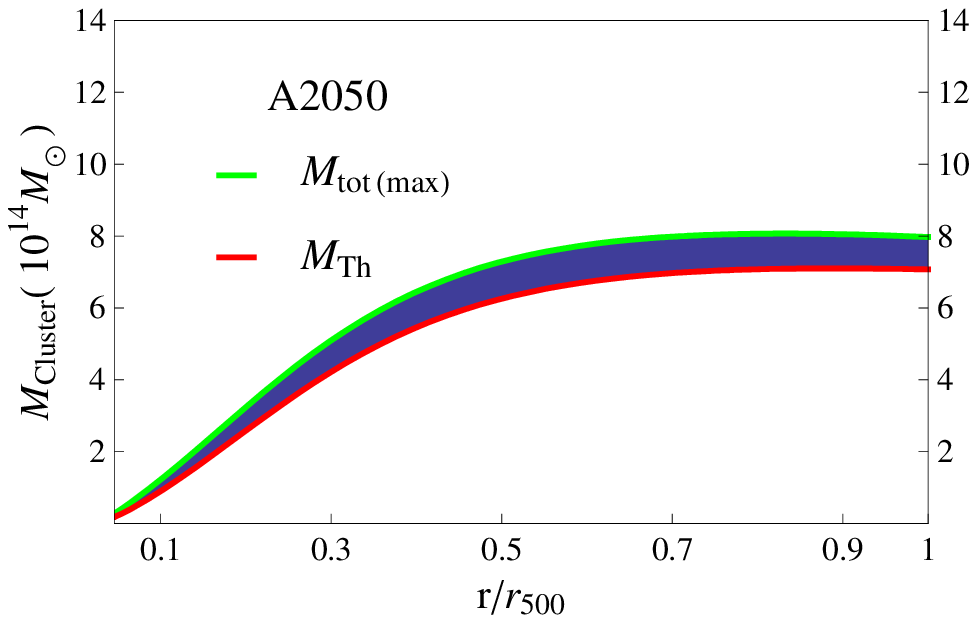}
\includegraphics[width=0.34\textwidth]{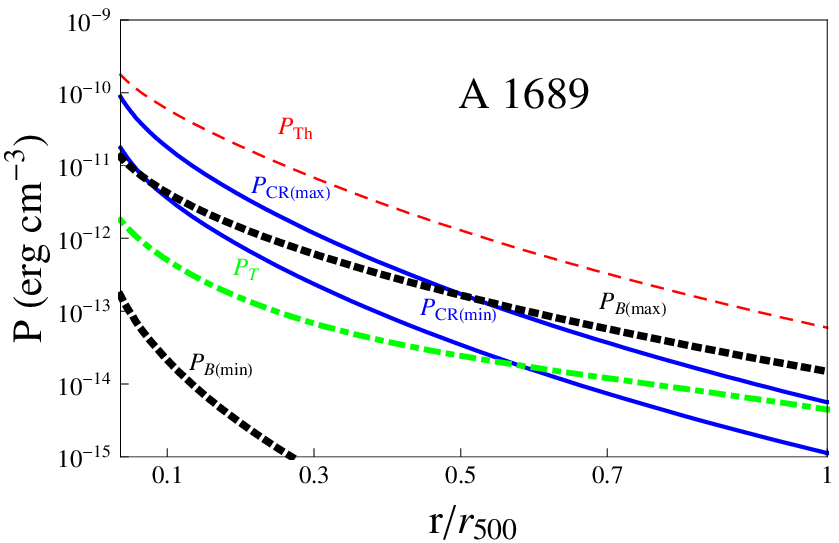}
\includegraphics[width=0.34\textwidth]{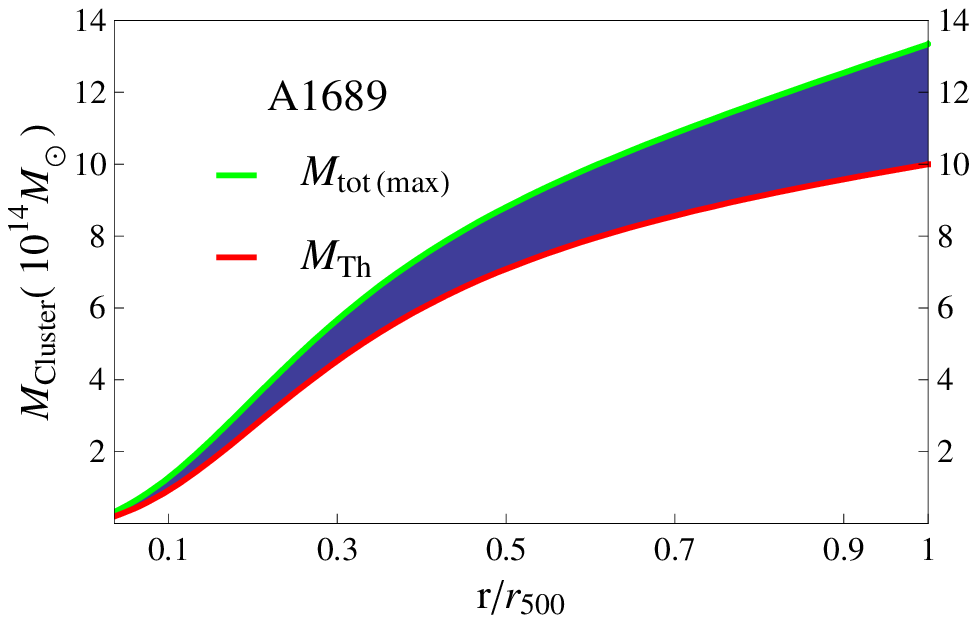}
\includegraphics[width=0.34\textwidth]{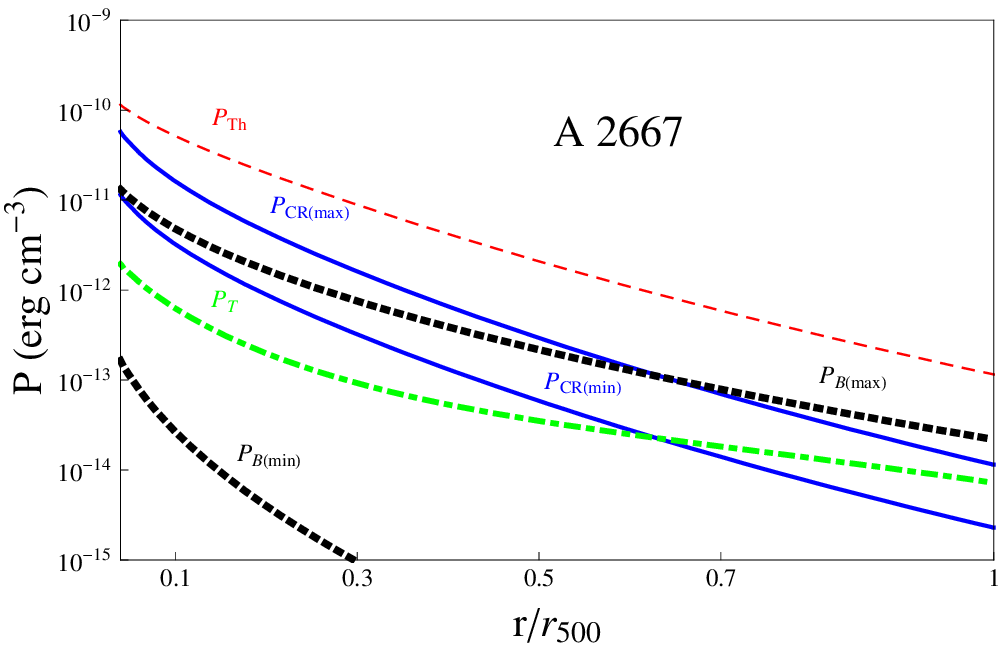}
\includegraphics[width=0.34\textwidth]{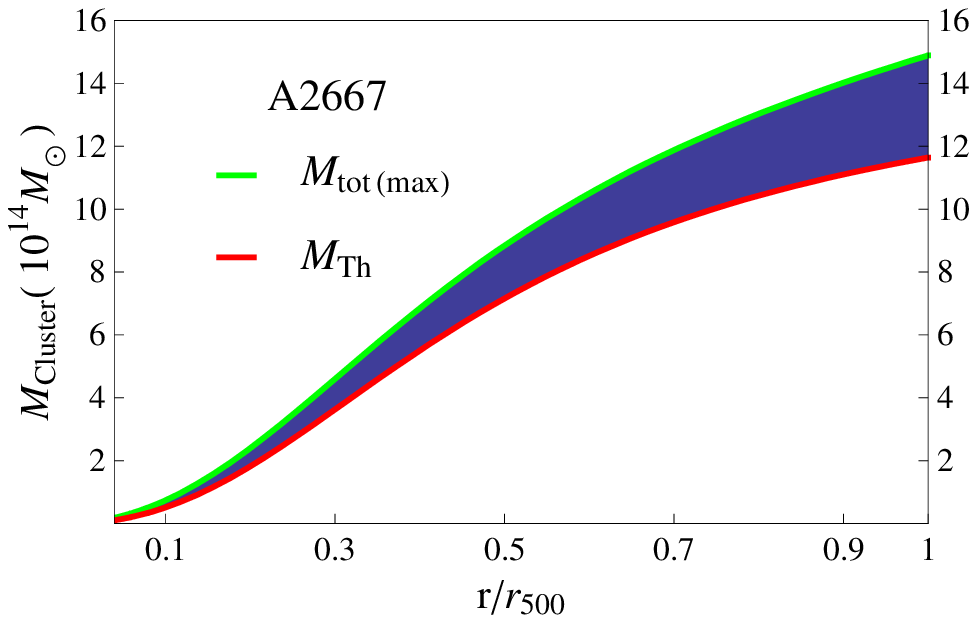}
\includegraphics[width=0.34\textwidth]{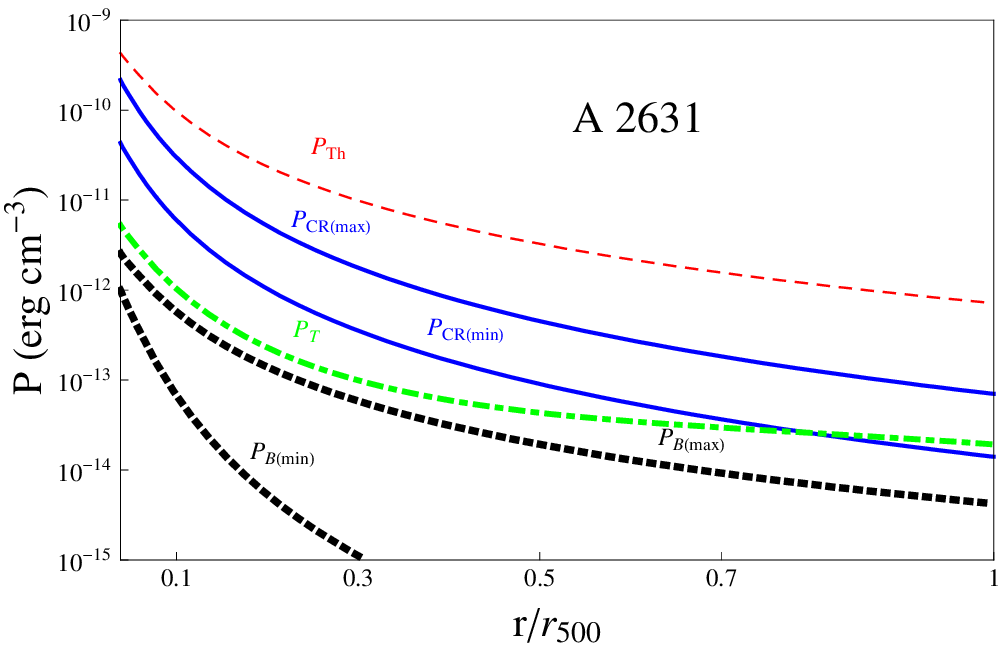}
\includegraphics[width=0.34\textwidth]{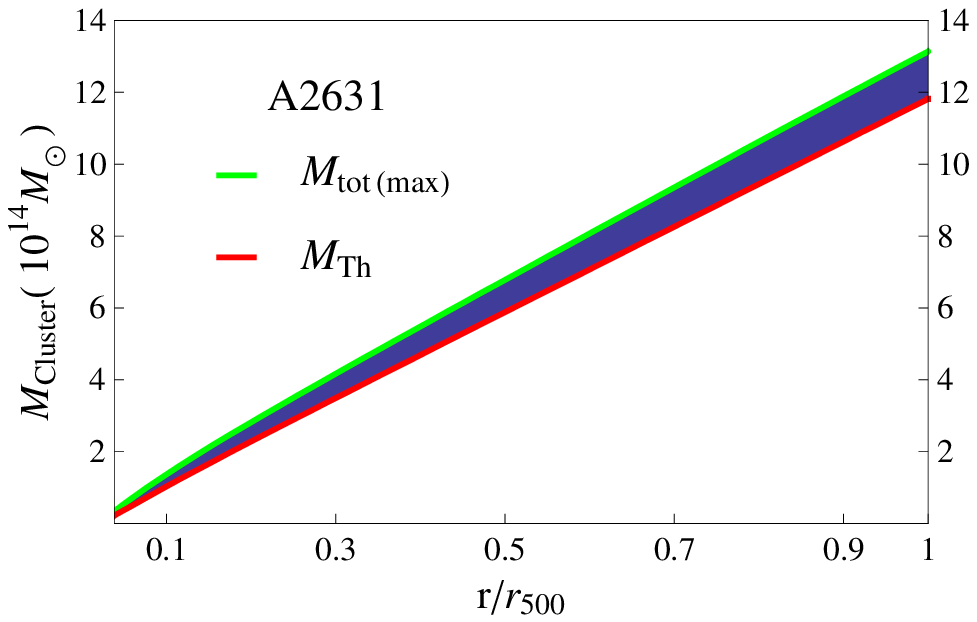}
\caption{\small Left panels: Minimum and maximum non-thermal pressure profiles for each component. The red dashed line 
represents the thermal pressure profile. For the non-thermal pressures, the blue continuous lines represent the 
maximum and the minimum cosmic ray pressure profiles, the thick black dashed lines represent the maximum and minimum magnetic
pressure profiles and the dot-dashed green line represents the turbulent pressure profiles. Right panels: comparison between
the hydrostatic mass profile (red lines) and the maximum mass profile due to non-thermal components. The blue region
represents all mass profile variations due to all the combinations possible for the three non-thermal components.
The clusters are displayed from top to bottom in the following sequence: A496, A2050, A1689, A2667 and A2631.}
\label{Pressures}
\end{figure*}

In Fig.~\ref{Pressures} we show the evaluation of all non-thermal pressures separately in order 
to analyze each contribution alone (right panels). The maximum influence which 
non-thermal components yield in the mass estimate is shown in left panels. 
From this Figure, we see that the main non-thermal contribution
comes from magnetic fields or cosmic rays, depending on the range of parameters adopted.

From this Figure we also see that in the inner parts ($r < \sim$0.5 $r_{500}$) of cool core clusters (A496, A1689 and A2667) the maximum
cosmic ray pressure is higher than the maximum magnetic pressure. On the other hand, 
for the outer parts of the mass profiles
the maximum magnetic pressure is always higher than the maximum cosmic ray pressure. 
However, this statement is not true for 
non-cool core clusters.
For A2050 and A2631, the cosmic ray pressure dominates the magnetic pressure for all radii.
Even the turbulent pressure becomes more important than the magnetic component 
(see Table~\ref{TabSigmaMass}).
It is important to note that we used results from the literature to estimate the
central magnetic field intensity \citep[e.g.,][]{Taylor93,Feretti99,Allen01,Eilek02} 
and as observed by \citet{Vogt05} the strength of central magnetic fields
in non-cool core clusters is lower than those present in cool core clusters.
Thus, as the magnetic pressure is described by Eq.~(\ref{eqB}), a lower central value
leads to a decrease in the magnetic pressure.

In Table~\ref{TabSigmaMass}, we present the maximum difference in mass estimates 
considering the influence of each non-thermal pressure separately and all three components together.
We define  $\sigma_{B}(\rm max)$ as the maximum difference
in mass estimates due to the magnetic pressure only, $\sigma_{\rm turb} (\rm max)$ is the maximum difference
in mass estimates due to the turbulence pressure only, $\sigma_{cr} (\rm max)$ maximum difference
in mass estimates due to the cosmic ray pressure only and $\sigma_{\rm total} (\rm max)$ is the 
maximum difference in mass estimates due to all non-thermal components.
The mass variation given by $\sigma M_{\rm NTP (\rm B)}$ is simply
\begin{equation}
\sigma M_{\rm NTP} = \frac{M_{\rm NTP} (r) - M (r)}{M (r)}.
\end{equation}

\begin{table}[ht]
\centering
\caption{Maximum difference in mass estimates.}
\begin{tabular}{ccccc}
\hline\hline
Cluster & $\sigma_{B}(\rm max)$  &   $\sigma_{\rm turb} (\rm max)$  &   $\sigma_{cr} (\rm max)$  &  $\sigma_{\rm total} (\rm max)$ \\
\hline
A496  & 17.33\% &  5.52\% &  4.87\%  & 27.72\%  \\
A1689 & 20.07\% &  3.87\% &  9.47\%  & 33.40\% \\
A2050 &  1.24\% &  0.82\% & 10.69\%  & 12.74\% \\
A2631 &  0.59\% &  0.77\% &  9.79\%  & 11.15\% \\
A2667 & 14.92\% &  3.06\% &  9.93\%  & 27.90\%  \\
\hline
\end{tabular}
\label{TabSigmaMass}
\end{table}

From Table~\ref{TabSigmaMass}, we see that for CC clusters the magnetic pressure is dominant, 
contributing more than 50\% of the total mass variation. 
Still, for NCC clusters the major component is the 
cosmic ray pressure, accounting for more than 80\% of the total mass variation.
We conclude that the maximum influence of non-thermal components on the total mass variation
ranges from more than 10\% to almost 35\%.
Although all of the assumptions agree with previous works, we emphasize that the difference 
in the hydrostatic mass estimates rely on the assumption of the specific parametrization of 
non-thermal components that were inferred based on numerical simulations and observational results.
As the errors in mass estimates (by weak lensing or via hydrostatic equilibrium) are in most cases
lower than the variation in mass due to non-thermal pressure contribution, we cannot neglect
this component.
Moreover, as X-ray data are widely used to constrain cosmological parameters, this evaluation 
should be regarded with care. Since in most cases this method neglects the non-thermal contribution to
mass estimates, these constraints may be biased low and consequently it will bias the cosmological determinations.
Thus, multi-wavelength study will play an important hole in the investigation of non-thermal components
in galaxy clusters.

\subsection{Constraining the non-thermal pressure based on A1689}

The plasma in many apparently relaxed systems may be affected by additional non-equilibrium
processes, which contribute to rise the pressure and hence cause an underestimate
of the cluster mass from X-ray observations of the thermal bremsstrahlung emission.
If this difference is due to non-thermal pressure, 
the comparison between independent methods of mass estimates can provide a
powerful constraint to the contribution of this component.

It is important to notice that weak lensing study is only possible if the cluster is massive and not very close. 
For the sample chosen in this work, the only cluster with available weak-lensing mass determination inside $r_{500}$ is A1689. 
In this section we focus on the comparison of a hydrostatic X-ray mass estimate determined in a previous work \citep{Lagana08}
with those derived from weak lensing which are available in the literature.
However, as noted by \citet{Hoekstra07}, it is difficult to compare results from 
different mass indicators since it would 
be necessary to make assumptions regarding the cluster geometry in the X-ray determination, 
and because lensing is sensitive to all mass along the line of sight, this would also bias weak 
lensing mass determination.

The total mass inside $r_{500}$ computed from X-ray measurements is $M_{500}=(11.14 \pm 0.46) \times 10^{14}  M_{\odot}$ 
without considering the non-thermal component.
For A1689, \citet{Mahdavi08} estimated a total mass inside $r_{500}$ of $M_{500}=(14.29 \pm 2.40) \times 10^{14}  M_{\odot}$.
In this case, the total non-thermal contribution (this means all three components: cosmic ray, magnetic and turbulent pressure)
to the mass estimates can range from $\sim$2\% to more than 30\%. 

This cluster is possibly undergoing a merger, where a sub-clump close to or along 
the line-of-sight is being accreted \citep{Andersson04}. Thus, in this specific case,
the turbulent motion seems to account for the most part of the non-thermal pressure. 
\citet{Nagai07} affirmed that the hydrostatic estimate of the total
mass is biased low by about 5\%-20\% through the virial region, 
primarily due to additional pressure support provided by subsonic 
bulk motions in the ICM.

On average, the hydrostatic cluster mass estimates are biased low by about 7.5\% at $r_{500}$
for relaxed systems, while the bias in unrelaxed systems is about 10.5\% at this radius \citep{Lau09}.
From our results (see Fig.~\ref{Pressures} for this cluster) we see that if one assume 
the non-thermal contribution for this cluster,
the mass estimates from different methods (e.g., weak lensing and X-rays) can be totally compatible.

\section{Conclusion}
\label{conc}
We have taken into account the effects of non-thermal pressure on the X-ray mass estimates for five Abell clusters
(A496, A2050, A1689, A2667 and A2631).
The masses derived considering just the thermal pressure were presented in a previous work by \citet{Lagana08}
and were used here for comparison.
We summarize our main results below:

\begin{itemize}

\item{The inclusion of non-thermal pressure in the intra-cluster gas description is
motivated by the increasing evidence for the presence of a magnetic field in clusters of galaxies.
We assume a magnetic profile given by $B(r) \propto B_{0} \rho_{g}^{\alpha}$, considering values for $B_{0}$
ranging from 5 up to 30 $\mu$G for CC clusters, while for NCC clusters we assume 2$\mu$G $< \rm B_{0}< $ 8$\mu$G. 
For each central value we let the shape parameter vary between 0.5 $ < \alpha <$ 0.9.
The magnetic pressure contributes with approximately 20\% to the total mass variation.}

\item{In order to take into account the influence of turbulent motion in the
ICM and bulk velocities, we assume isotropic turbulent pressure based
on the numerical simulation results of \citet{Lau09}: 
$P_{\rm turb}=1/3 \rho_{\rm g} (\sigma_{r}^{2}+\sigma_{\rm t}^{2})$.
The tangential ($\sigma_{r}$) and the radial ($\sigma_{\rm t}$) dispersion velocity profiles 
were derived based in the same numerical simulations. This component can influence up to 5\%
of the total mass estimates.}

\item{Energetic particles are confined by magnetic fields. Since ICM is permeated by
magnetic fields, cosmic rays can also provide an important source of pressure.
As the distribution of cosmic rays is poorly known, we considered the prescription
of \citet{Ando08} to derive the cosmic ray pressure. From our results we can see that
this component can affect the cluster mass estimates by $\sim$ 10\%. In the inner parts of 
cool core clusters this component is comparable to the magnetic pressure.}

\item{The plasma in clusters of galaxies may be affected by non-thermal processes which rise
the pressure and hence cause an underestimate in X-ray measurements of the total mass.
In this way the comparison between independent methods of mass estimates 
can constrain the contribution of non-thermal
pressure. We compared weak-lensing results with X-ray mass 
measurements for A1689 to investigate this, and if the difference in mass estimates is in fact due 
to the non-thermal component, it can account
for 2\% to $\sim$30\%.}

\item{We took into account the effects of non-thermal pressure on the total mass estimates.
To derive the thermal pressure and compute the hydrostatic mass we used observed 
temperature and density profiles \citep{Lagana08}. 
This is the first study that considers the influence of all three
non-thermal pressure on the total mass estimates. 
Within the limits of our sample and with several reasonable assumptions 
to describe each of the non-thermal components, it is important to notice that our 
findings rely on the specific parametrization adopted in this work.
This study indicates that further investigations
are needed to make a detailed description of the influence of non-thermal components in ICM.}

\item{Without the complete knowledge of the non-thermal contribution, 
we should be aware of the fact that clusters of galaxies have been used as 
observational tools for investigations of cosmological interest based only on the hydrostatic
assumption. The non-thermal components are neglected. 
A thorough knowledge of non-thermal contribution will come from the 
combination of X-ray, radio and gamma ray data for a large sample of clusters.
The next generation of cluster surveys will provide data
to address these fundamental questions in cosmology\footnote{http://wfxt.pha.jhu.edu/}.}

\end{itemize}

\begin{acknowledgements}
We acknowledge Gast\~{a}o B. Lima Neto, Renato A. Dupke and Elisabete M. De Gouveia Dal Pino for stimulating discussions. 
We also acknowledge the anonymous referee for the stimulating suggestions which improved our work.
We gratefully thank Florence Durret, Reuven Opher and Emille Ishida for revising this manuscript.
This work was support by the Brazilian agency FAPESP (grants: 08/04318-7, 04/05961-0, 06/58240-3).
\end{acknowledgements}

\bibliographystyle{aa}
\bibliography{lsk09_nov}

\end{document}